\documentclass[aps,prx,preprint,groupedaddress,amssymb,amsmath,floatfix]{revtex4-1} 

\usepackage{graphicx}
\usepackage{bm}
\usepackage{cleveref}

\newcommand{\moire}{moir\'e\:}
\newcommand{\Moire}{Moir\'e\:}
\newcommand{\WS}{WS$_{2}$\:}
\newcommand{\MoSe}{MoSe$_{2}$\:}

\newcommand{\degree}{$^{\circ}$}

\usepackage{subfig}
\usepackage{color}

\begin{document}
\title{\Moire lattice-induced formation and tuning of hybrid dipolar excitons in twisted WS$_{2}$/MoSe$_{2}$ heterobilayers}
\author{Long Zhang$^1$, Zhe Zhang$^{1,2}$, Fengcheng Wu$^3$, Danqing Wang$^1$, Rahul Gogna$^4$, Shaocong Hou$^5$, Kenji Watanabe$^6$, Takashi Taniguchi$^6$, Krishnamurthy Kulkarni$^5$, Thomas Kuo$^1$, Stephen R. Forrest$^{1,5}$}
\author{Hui~Deng$^{1,4}$} \email{dengh@umich.edu}
\small{ }
\address{$^1$ Physics Department, University of Michigan, 450 Church Street, Ann Arbor, MI 48109-2122, USA}
\address{$^2$ State Key Laboratory of Surface Physics, Department of Physics, Fudan University, Shanghai 200433,China}
\address{$^3$ Condensed Matter Theory Center and Joint Quantum Institute, Department of Physics, University of Maryland, College Park, Maryland 20742, USA}
\address{$^4$ Applied Physics Program, University of Michigan, 450 Church Street, Ann Arbor, MI 48109-1040, USA}
\address{$^5$ Department of Electrical Engineering and Computer Science, University of Michigan, 450 Church Street, Ann Arbor, MI 48109-1040, USA}
\address{$^6$ National Institute for Materials Science, Tsukuba, Japan}


\begin{abstract}
\Moire superlattices formed in van der Waals bilayers have enabled the creation and manipulation of new quantum states, as is exemplified by the discovery of superconducting and correlated insulating states in twisted bilayer graphene near the magic angle. Twisted bilayer semiconductors may lead to tunable exciton lattices and topological states, yet signatures of moir\'e  excitons have been reported only in closely angularly-aligned bilayers. Here we report tuning of \moire lattice in \WS/\MoSe bilayers over a wide range of twist angles, leading to the continuous tuning of \moire-lattice induced interlayer excitons and their hybridization with optically bright intralayer excitons. A pronounced revival of the hybrid excitons takes place near commensurate twist angles, $21.8^\circ$ and $38.2^\circ$, due to interlayer tunneling between states connected by a \moire reciprocal lattice vector. From the angle dependence, we obtain the effective mass of the interlayer excitons and the electron inter-layer tunneling strength. These findings pave the way for understanding and engineering rich \moire-lattice induced phenomena in angle-twisted semiconductor van dar Waals heterostructures.
\end{abstract}


\maketitle
The ability to create atomically thin heterostructures by stacking van der Waals materials marks a new frontier in materials science and condensed matter physics\cite{geim_Van_2013,novoselov_2D_2016,rivera_valley-polarized_2016}. When two monolayer crystals of the same lattice symmetries overlay on each other, a moir\'e superlattice may form due to a small mismatch in their lattice constants or angular alignment \cite{kuwabara_Anomalous_1990,zhang_interlayer_2017}. The latter -- the twist angle between the two layers -- provides a unique tuning knob for engineering the electronic properties of the heterostructure. Seminal results have been obtained in twisted bilayer graphene, where superconducting and correlated insulating states are created by fine control of the twist angle \cite{cao_Unconventional_2018,cao_correlated_2018,chen_Signatures_2019,chen_Evidence_2019}.
In semiconductors, such as transition metal dichalcogenides heterobilayers, a wide variety of phenomena become possible and tunable with angle in \moire lattices, ranging from single exciton arrays to topological bands, to strongly correlated states \cite{wu_Topological_2017,yu_Moire_2017,wu_theory_2018,wu_Hubbard_2018}. 
Localized excitons, exciton mini-bands, and correlated states in \moire superlattices have been reported in TMD bilayers with very small twist angles \cite{jin_Observation_2019,tran_Evidence_2019,seyler_Signatures_2019,wang_Evidence_2019,tang_wse2/ws2_2019,regan_optical_2019,shimazaki_moire_2019}. Increasing the twist angle, however, has led to suppression of measurable effects of \moire superlattices. Hybrid exciton resonances formed by coupling between inter- and intra-layer excitons have also been reported in angularly-aligned \WS/\MoSe heterobilayers and it was suggested that the \moire lattice may enhance interlayer coupling; yet only a single resonance was resolved as the twist angle deviates from $0^\circ$ or $60^\circ$ \cite{alexeev_Resonantly_2019}. 

In this work, we demonstrate twist-angle tuning of \moire lattices in \WS/\MoSe bilayers and show how the \moire lattices drastically change the properties of excitons.
Due to band folding in the \moire reciprocal lattice and electron tunneling in the \moire lattice, new, momentum-direct interlayer exciton states are formed which can hybridize with nearly-resonant intralayer excitons, leading to two bright hybrid \moire exciton states
over a wide range of  twist angles $\theta$
.  
The role of the \moire lattice is corroborated by revival of strong inter-layer tunneling when $\theta$ is near $21.8^\circ$ and $38.2^\circ$. At these angles, \moire-lattices are formed commensurate with the monolayer lattices, bringing angularly shifted valleys of the two monolayers into equivalent momentum in the \moire Brillouin zone, thereby enabling strong inter-layer tunneling. The resulting hybrid exciton states resemble the features in hetero-bilayers with $\theta =60^\circ$ and $0^\circ$, respectively.
The observation of these new \moire states in the twisted bilayers directly demonstrates the discrete translational symmetry, or periodicity, of the \moire pattern --- the defining feature of a lattice.

We furthermore measure continuous tuning of the energy and spectral weight of the hybrid excitons in the \moire lattice by the twist angle. The results agree very well with an analytical theory model that takes into account the interplay between spin-orbit splitting of the conduction band, valley selection rules, atomic stacking orders and the lattice symmetries.
By comparing with the model, we obtain from the angle dependence the effective mass of the interlayer excitons and the inter-layer electron tunnelling strength.
Lastly, by comparison of hybrid states formed from different intra-layer exciton states, we find that the binding energies of the uncoupled inter-layer and intra-layer exciton states differ by merely 10-16~meV, as opposed to $\sim$100~meV based on first-principle calculations when interlayer tunneling is neglected \cite{latini_Interlayer_2017}.

\begin{figure*}[t]
	\includegraphics[width=\linewidth]{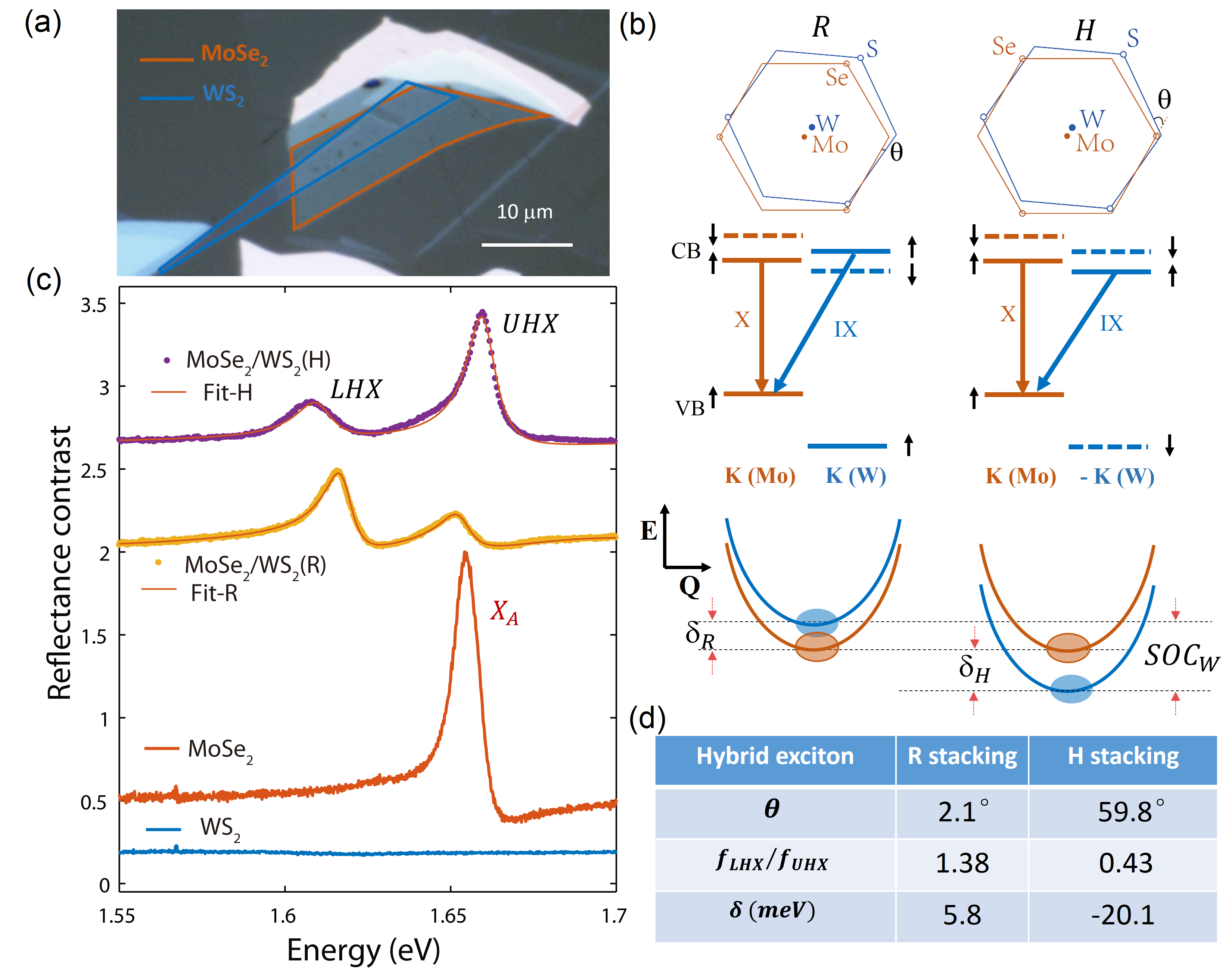}
	\caption{Hybrid excitons in rotationally aligned \WS/\MoSe bilayers. (a) Optical microscope image of an hBN-encapsulated heterobilayer. Red and blue solid lines outline the \MoSe and \WS monolayers, respectively. (b) Top panels illustrate a unit cell of R-stacking (left) and H-stacking (right) \WS/\MoSe bilayers with a small twist angle. Middle panels depict the corresponding conduction and valence band alignment of \WS and \MoSe, where X labels the intra-layer transition and IX labels the nearly-resonant inter-layer transition that shares the same hole state. Solid and dashed lines corresponds to states of opposite spins. Bottom panels illustrate the alignment between an intra-layer \MoSe A exciton state (red) and the inter-layer exciton state (blue) that it hybridizes with. (c) Reflection contrast (RC) spectra for, from bottom to top, a monolayer \WS (blue), monolayer \MoSe (red), R-stacking bilayer (orange) and H-stacking bilayer (purple). The dots are data and solid lines are fits by transfer matrix calculations. (d) Summary of the ratio of the oscillator strength between LXH and UHX by fitting the bilayer spectra in c, and the corresponding detuning $\delta$ calculated using the coupled oscillator model.}
    \label{fig:band}
\end{figure*}

The devices used in this work are \WS/\MoSe heterobilayers with different twist angles $\theta$, encapsulated by few-layer hexagonal boron nitride (hBN) (see Methods for details). Figure~\ref{fig:band}a shows an example of the optical microscope image of an encapsulated heterobilayer, where the sharp edges of two monolayers are aligned. The twist angle is measured to be $\theta=59.8^\circ\pm 0.3^\circ$ by comparing the angle-dependence of second harmonic generation from each monolayer and from the bilayer (see Supplementary Information SI figure2 for details).
The hole band offset is large but the conduction band offset is small between \WS and \MoSe. Therefore inter-layer electron tunneling is expected between states of the same spin and valley, which leads to hybridization between the corresponding intra- and inter-layer exciton transitions that share the same hole state, as illustrated in the inset of Figure~\ref{fig:band}a.

We first characterize exciton hybridization in closely aligned hetero-bilayers, with $\theta\sim 0^\circ$ or $60^\circ$. In such bilayers, the Brillouin zones of the two layers closely overlap in momentum space to form nearly direct bandgaps for both the inter- and intra-layer transitions (Figure~\ref{fig:band}b), and thereby inter- and intra-layer excitons can hybridize \cite{hsu_Tailoring_a}.

To identify optically bright resonances, we measure the reflectance contrast (RC) spectrum, $\frac{R_{sample}-R_{substrate}}{R_{substrate}}$ (see Methods for details). Interlayer exciton has an oscillator strength two orders of magnitude weaker than that of the intra-layer exciton, due to separation of the electron and hole wavefunction, so it is typically too weak to be measurable in absorption or RC spectroscopy. However, when interlayer excitons hybridize with intra-layer ones via electron or hole tunneling,  the hybrid states acquire an oscillator strength through the intra-layer exciton component. Therefore, we can identify the hybrid excitons via their spectral weight in the absorption spectra of the heterobilayer in comparison to those of the monolayer.

As shown in \cref{fig:band}c, the \MoSe monolayer region of the device (as marked on \cref{fig:band}a) shows a strong intralayer \MoSe A exciton resonance near 1.65~eV, while the \WS monolayer has no exciton resonances nearby. In the bilayer, stacking of the \WS layer is expected to lead to a red shift of the \MoSe A exciton resonance\cite{alexeev_Resonantly_2019} while also introducing an interlayer exciton transition, between an electron in \WS and a hole in \MoSe, that is close in energy. The interlayer exciton has a negligible oscillator strength due to electron-hole separation. However, two clearly-resolved resonances appear in our bilayer, both with significant spectral weight (top spectrum in \cref{fig:band}c). The same two resonances are also measured in photoluminescence (See supplementary material SI figure 3). We therefore identify them as the inter- and intra-layer hybrid states, both of which inherit an oscillator strength from their intra-layer component \cite{alexeev_Resonantly_2019}.
There are multiple pairs of intra- and inter-layer excitons that can hybridize. We focus on the transition region of \MoSe A exciton first and label these states as $MoA$ excitons, of which the hole is always in the highest \MoSe valence band. Other pairs will be analyzed later.

The relative spectral weight of the lower (LHX) and upper hybrid excitons (UHX) corresponds to the ratio of their oscillator strengths $f_{LHX}/f_{UHX}$. It is controlled by their intra-layer exciton fractions, which in turns is controlled by the energy detuning  $\delta=E_{IX}-E_{X}$ between the uncoupled inter-layer ($E_{IX}$) and intra-layer ($E_{X}$) resonances. Therefore $f_{LHX}/f_{UHX}$ greater or less than one corresponds to positive or negative detuning $\delta$.

As clearly seen in the spectra (\cref{fig:band}c), in the R-stacking bilayers ($\theta=2.1^{\circ}$), the LHX has a larger spectral weight than the UHX, suggesting the uncoupled interlayer state lies above the intralayer one, or $\delta_R>0$, as illustrated in \cref{fig:band}b. In contrast, in the H-stacking bilayer ($\theta=59.8^{\circ}$), the UHX has a larger spectral weight, suggesting  $\delta_H<0$. These results are consistent with the spin selection rules of the excitonic transitions (see more detailed illustration in Fig.~\ref{fig:band}b) \cite{hsu_Tailoring_a}. The difference in the detuning between R- and H-stacking bilayers correspond to the spin-orbit splitting of \WS conduction band (see illustration in \cref{fig:band}b).

To analyze the results quantitatively, we first obtain the energies, $E_{LHX}$ and $E_{UHX}$, and oscillator strengths of the hybrid states by fitting the RC spectra using the transfer matrix method, where the hybrid excitons are modeled as Lorentz oscillators (See supplementary information SI figure1 for details) \cite{hsu_Tailoring_a,li_Measurement_2014}. The fitted spectrum agrees well with the data, as shown in \cref{fig:band}c.  Describing the hybrid modes with the coupled oscillator model, we have $E_{LHX}-E_{UHX}=-\sqrt{4J^2+\delta^2}$, and$\frac{f_{LHX}}{f_{UHX}} = \frac{\sqrt{\delta^2+4J^2}+\delta}{\sqrt{\delta^2+4J^2}-\delta}$ (see Methods for details). Thereby using the fitted $E_{LHX, UHX}$ and $f_{LHX, UHX}$, we can obtain $\delta$ and $J$. The results are summarized in \cref{fig:band}d. The coupling strength is found to be around 20~meV for both R- and H-stacking.  The detuning of the R- and H-stacked bilayers differ by $25.9\pm0.5$~meV \footnote{Considering the lattice mismatch between the two layers (see Methods) would add a small correction to the measured spin-orbit splitting. Using Eq.~1 and the fitted effective masses, we obtain a value of 26.4 meV for spin-orbit coupling based on the measurement.}, consistent with the spin-orbit splitting of \WS from theory calculation \cite{liu_Electronic_2015}, confirming the hybrid states are formed by spin-conserved electron tunneling between the two layers. 



To study tuning of the hybrid excitons by the \moire lattice, we perform the same measurements and analysis as discussed above on 30 samples with different twist angles, and obtain how the exciton energies, oscillator strengths and inter-layer tunneling vary with the changing \moire lattice. We define $\theta_0 < 30^\circ$ as the angular deviation from aligned bilayers of R- or H-stacking. $\theta_0 = |\theta|$ for R-stacking and $\theta_0 = |60^{\circ}-\theta|$ for H-stacking.  

\begin{figure*}[t]
  \includegraphics[width=\linewidth]{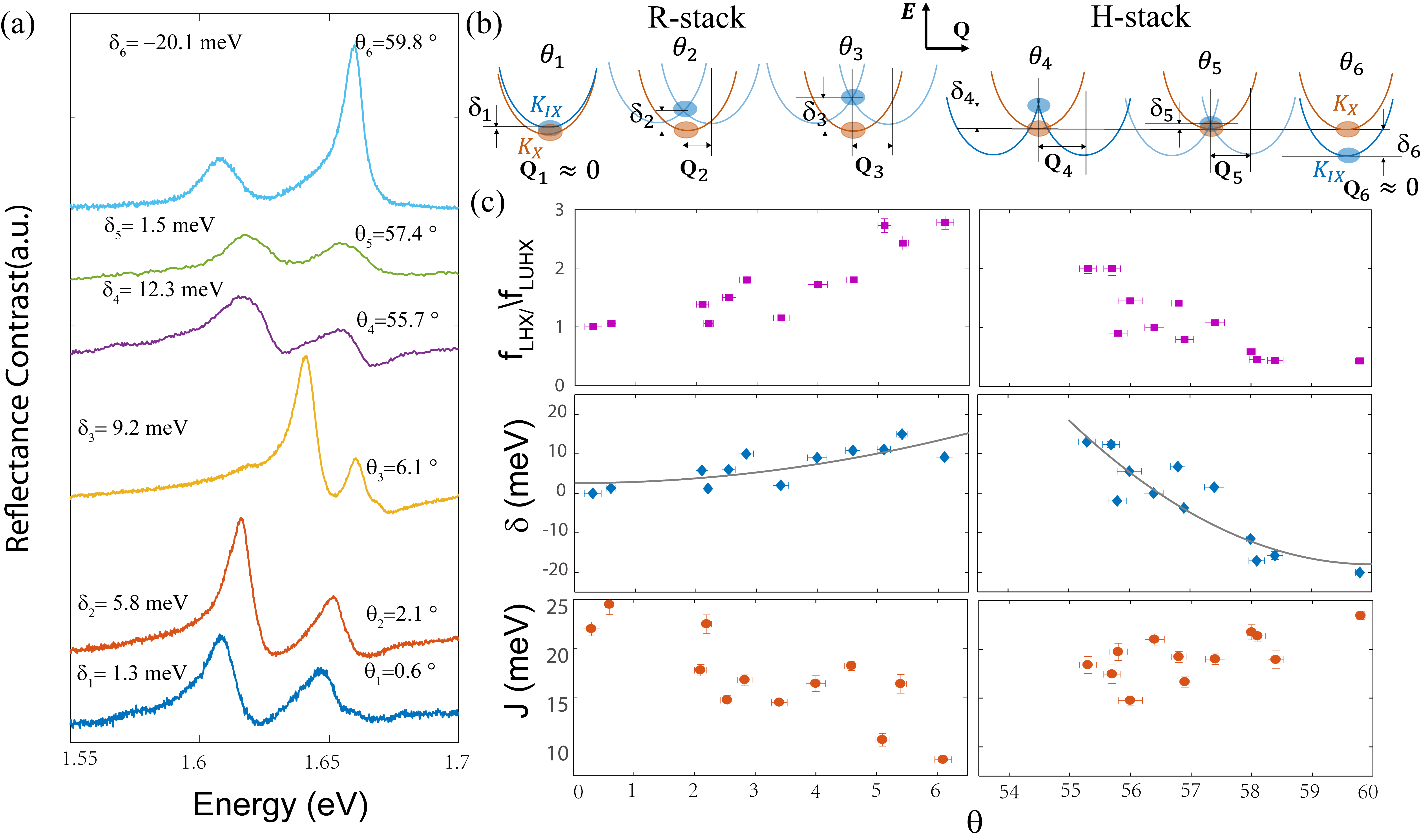}
	\caption{Twist angle dependence of the hybrid excitons. (a) RC spectra of bilayers of different twist angles $\theta_i$, for $i=1$ to $6$. The corresponding $\theta_i$ and fitted detuning $\delta_i=E_{IX,i}-E_{X,i}$ are labeled by each spectrum. Two well-resolved, bright hybrid $MoA$ excitons are observed for $\theta_0$ up to $6^\circ$. (b) Schematics of $MoA$ intra-layer (red) and inter-layer (blue) exciton bands at the different twist angles $\theta_i$. The interlayer exciton band is displaced in the momentum space by $Q_i$ with increasing $\theta_i$. The moir\'e superlattice leads to band folding and formation of a new interlayer exciton state at the $\Gamma$ point $q=0$ (blue oval), with the same angular momentum as the intralayer exciton state (red oval). (c) Ratio of the oscillator strengths of the hybrid states $LHX_{MoA}$ and $UHX_{MoA}$, detuning, and inter- and intra-layer exciton coupling strength as a function of the twist angle $\theta$, obtained from the RC spectra. The gray solid lines in the middle panel are quadratic fits based on equation~\ref{deltatheta}. 
       }
	\label{fig:twist}
\end{figure*}

As shown in \cref{fig:twist}, the $MoA$ hybrid exciton doublets are clearly resolved for $\theta_0$ up to $6^\circ$, which would correspond to a tuning of the \moire lattice constant by nearly three-fold \cite{rasmussen_Computational_2015}.  The spectral weights of the doublets evolve continuously with the twist angle, 
reflecting continuous increase of $f_{LHX}/f_{UHX}$ and $\delta$ with $\theta_0$ (middle panel of \cref{fig:twist}c). At the same time, the inter-layer tunneling $J$ decreases continuously (bottom panel of \cref{fig:twist}c). These observations show clearly \moire lattice induced hybridization and tuning of intra-layer and inter-layer excitons, as we explain below.

We illustrate in \cref{fig:twist}b the $MoA$ exciton bands at different twist angles, corresponding to the six samples shown in \cref{fig:twist}a.
The intra-layer \MoSe A exciton transition (red band) remains direct, with the band minimum at zero center-of-mass momentum $\boldsymbol{q}_X\sim 0$, irrespective of the twist angle.
It is close in energy with the inter-layer exciton formed by a hole from the same \MoSe valence band but an electron from a \WS conduction band. This inter-layer exciton band has the band minimum also at zero center of mass momentum: $\boldsymbol{q}_{IX}\sim 0$, when $\theta \sim 0^\circ$ ($\theta_1$ in \cref{fig:twist}a-b)  or $60^\circ$ ($\theta_6$ in \cref{fig:twist}a-b), neglecting the small lattice constant mismatch. This is the situation discussed in \cref{fig:band}.

As the two lattices rotate relative to each other by $\theta$ ($\theta_2$ to $\theta_5$  in \cref{fig:twist}a-b), the Brillouin zones of \MoSe and \WS also rotate by $\theta$.
The inter-layer exciton band minimum shifts away from the intra-layer exciton band minimum by momentum $\boldsymbol{K}_W-\boldsymbol{K}_M$ for R stacking, where $\boldsymbol{K}_M$ and $\boldsymbol{K}_W$ are respectively the Brillouin zone corners for \MoSe and \WS layers. Due to this momentum mismatch, hybridization between intra-layer \MoSe A excitons and the interlayer state at the band minimum is not allowed. 

\begin{figure*}[t]
  \includegraphics[width=\linewidth]{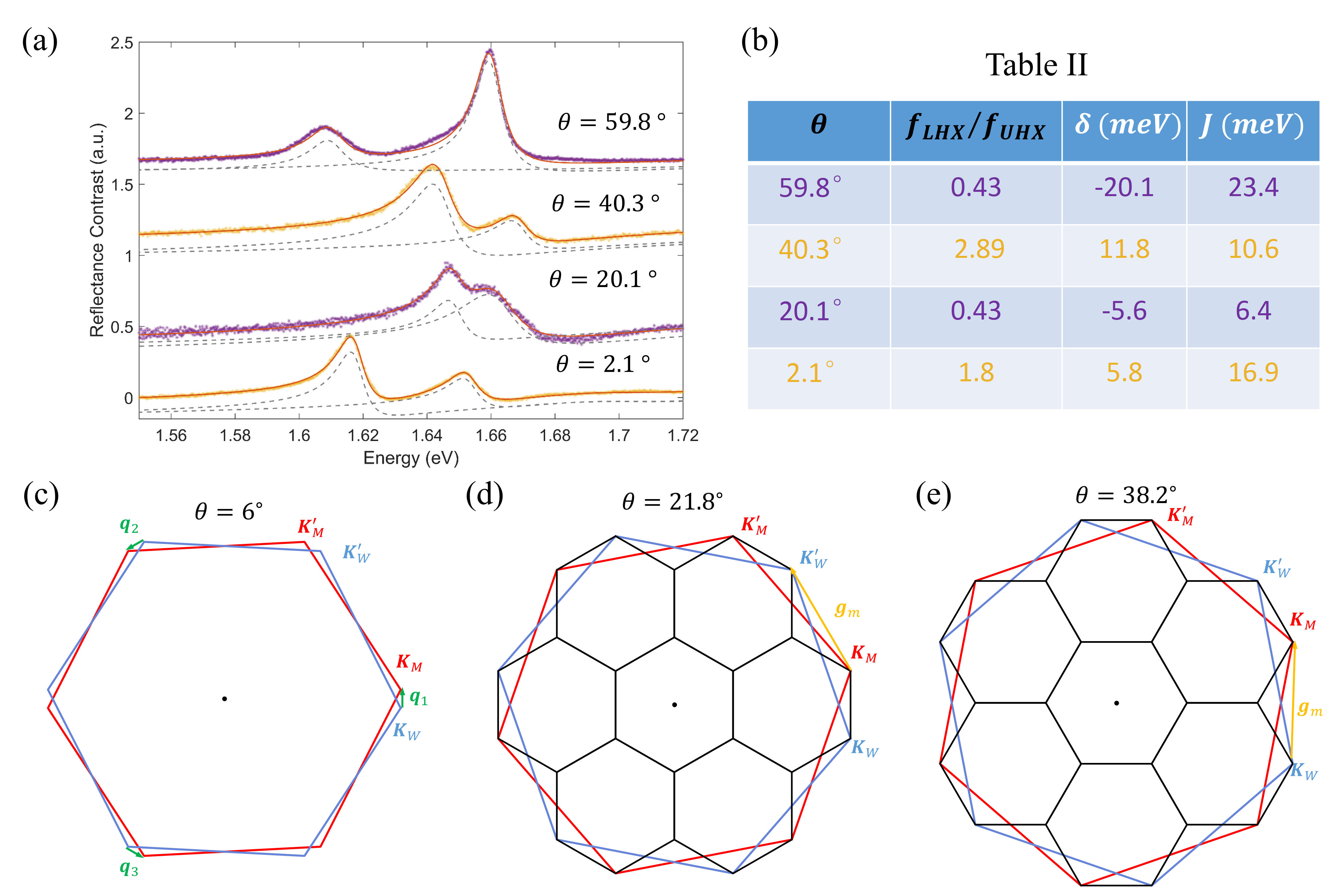}
	\caption{Hybridization in commensurate \moire lattices compared with aligned bilayers. (a) RC spectra of bilayers with $\theta=2.1^\circ , 20.1^{\circ}, 40.3^{\circ}$ and $59.8^{\circ}$. All show two well-resolved hybrid excitons. The LHX has a higher (lower) spectral weight than UHX has in bilayers with $\theta=2.1^\circ$ and $40.3^{\circ}$ ($\theta=59.8^\circ$ and $20.1^{\circ}$). 
Dots are data, solid lines are fits by transfer matrix calculation, and dashed lines are the fitted individual hybrid exciton resonances. (b) Summary of the fitted parameters for the RC spectra in a, showing similarities between bilayers with $\theta=2.1^\circ$ and $40.3^{\circ}$ and between bilayers with $\theta=59.8^\circ$ and $20.1^{\circ}$.  (c-e) Schematics of the Brillouin zones of twisted bilayers. The red (blue) hexagons depict the Brillouin zones of MoSe$_2$ (WS$_2$) monolayers. The lattice constant mismatch between MoSe$_2$ and WS$_2$ is neglected in this figure for the purpose of clear illustration. In (c), the twist angle is small, at $\theta=6^\circ$. The green arrows indicate vectors $\boldsymbol{q}_1$, $\boldsymbol{q}_2$ and $\boldsymbol{q}_3$, which represent the momentum shift between the Brillouin zone corners of the two monolayers.
In (d), $\theta=21.8^{\circ}$, a commensurate \moire lattice is formed, with the corresponding \moire Brillouin zone depicted by the black hexagons.
The yellow arrow represents the moir\'e reciprocal lattice base vector that connects $\boldsymbol{K}_M$ and $\boldsymbol{K}_{W}'$.
In (e), $\theta=38.2^{\circ}$, which is another commensurate angle dual to $21.8^{\circ}$, and $\boldsymbol{K}_M$ and $\boldsymbol{K}_{W}$ become equivalent states in the moir\'e Brillouin zone.
}
    \label{fig:coupling}
\end{figure*}

However, interlayer electron tunneling in the moir\'e lattice can lead to the formation of new moir\'e minband states to hybridize with the optically bright intralayer excitons. As illustrated in \cref{fig:twist}b and \cref{fig:coupling}c, three interlayer excitons $| \boldsymbol{q}_i \rangle_{\text{IX}} $ overlap with the optically bright intralayer exciton, where the center of mass momentum $\boldsymbol{q}_i$, measured relative to the band minimum of interlayer exciton, correspond to $\boldsymbol{q}_1 = \boldsymbol{K}_M-\boldsymbol{K}_W$ for $R$-stacking, with $\boldsymbol{q}_{2,3}$ connected to $\boldsymbol{q_1}$ by $2\pi/3$ and $4\pi/3$ rotations, respectively, via \moire reciprocal lattice vectors. These three inter-layer states are offset from their band minimum by the kinetic energy $\hbar^2 q_i^2/(2 M_{\text{IX}})$, for $i=1, 2, 3$ and $M_{\text{IX}}$ the  total mass of inter-layer exciton. 
These three states can couple due to the moir\'e lattice and therefore, superpose to form moir\'e minband states, of which one interlayer exciton state  shares the same angular momentum as the intra-layer \MoSe A exciton at $q_X \sim 0$, giving rise to the hybrid doublet we observe in angularly misaligned bilayers (see Methods for details).

When $\theta$ deviates more from $0^\circ$ or $60^\circ$, the interlayer exciton formed in the \moire lattice continuously blueshifts because of the increasing kinetic energy, which explains the measured continuous blueshift of the LHX and UHX resonances, and the continuous increase of the spectral weight of LHX compared to UHX.


To analyze our results more quantitatively, we develop an analytical microscopic theory based on the above understanding (see Methods for details). Comparing it with the measured twist-angle dependence of the hybrid states, we obtain the key band parameters of the bilayer, including the interlayer exciton effective mass and interlayer coupling strength.

We first compare the measured detuning $\delta$ vs. $\theta_0$ and the interlayer exciton kinetic energy. As discussed above, $\delta$ is given by:
\begin{equation}
\delta= \delta_0 + \frac{\hbar^2 \boldsymbol{q}_1^2}{2 M_{\text{IX}}},
\label{deltatheta}
\end{equation}
where $\delta_0$ is the detuning at $\theta = 0^\circ$ or $60^\circ$ for bilayers close to R- and H-stacking, respectively. $\boldsymbol{q}_1$ is equal to $4 \pi/ (3 a_M)$, and $a_M$ is the moir\'e period approximated by $a_0/\sqrt{\theta_0^2+\epsilon^2}$, for $a_0$ the monolayer lattice constant and $\epsilon$ the lattice constant mismatch $|a_0-a_0'|/a_0$ between the two layers. Equation.~(\ref{deltatheta}) shows that $\delta$ increases quadratically with $\theta_0$. As $\theta_0$ increases from 0\degree to 6\degree, $a_M$ changes by nearly three fold, and $\delta-\delta_0$ changes by seven-fold \cite{rasmussen_Computational_2015}.
Fitting the measured $\delta$ vs. $\theta_0$ with Equation.~(\ref{deltatheta}), we find the inter-layer exciton total mass $M_{\text{IX}}$ to be $(6.9 \pm 3.2) m_{0}$ and $(1.41\pm0.28) m_{0}$ for $R$- and $H$-stacking heterobilayers, respectively, for $m_0$ the electron free mass.

From our microscopic theory, we can also estimate the conduction-band interlayer tunneling parameter $w$ from the coupling strength $J$ through the relation
\begin{equation}
J= \frac{\sqrt{3} w} {\mathcal{A}} \sum_{\boldsymbol{k}} \phi_{\boldsymbol{k}+\frac{m_{h,IX}}{M_{\text{IX}}} \boldsymbol{q}_1}^* \psi_{\boldsymbol{k}},
\label{Jtheory}
\end{equation}
where $\phi_{\boldsymbol{k}}$ and $\psi_{\boldsymbol{k}}$ are respectively the relative-motion wave function for interlayer and intralayer excitons with the normalizations $(1/\mathcal{A})\sum_{\boldsymbol{k}} |\psi_{\boldsymbol{k}}|^2=1$ and $(1/\mathcal{A})\sum_{\boldsymbol{k}} |\phi_{\boldsymbol{k}}|^2=1$. Here $\mathcal{A}$ is the system area and $m_{h,IX}$ is the hole mass for the interlayer exciton.  Because of the momentum shift $(m_{h,IX} / M_{\text{IX}}) \boldsymbol{q}_1$ in the integral of Eq.~(\ref{Jtheory}),  $J$ decreases with increasing $\theta_0$, which agrees with the experimentally observed angle dependence of $J$. At small $\theta_0$, $J$ can be approximated by $\sqrt{3} w$. Using our experimentally measured value of $J$ at $\theta_0\sim 0$, we estimate the interlayer tunneling $w$ to be about $14$~meV for both R- and H-stacking bilayers. 

When the twist angle $\theta_0$ is greater than 6\degree, the hybrid exciton doublets become hard to be resolved, likely because there is a large blue detuning and the UHX has a vanishing oscillator strength (See supplementary material SI figure 4 for details). 

Remarkably, pronounced and well-resolved doublets re-appear in hetero-bilayers with $\theta =20.1^\circ\pm 0.3$ and  $40.3^\circ \pm0.3$, as shown in Fig.~\ref{fig:coupling}. In the bilayer with $20.1^\circ$ twist angle, the LHX has a smaller spectral weight than UHX has, corresponding to a negative detuning ($\delta = -5.6$ meV), which is similar to H-stacking bilayers formed at $\theta \sim 60^{\circ}$. In contrast, in the bilayer with $40.3^\circ$ twist angle, the LHX has a larger spectral weight than UHX has, corresponding to a positive detuning ($\delta=11.8$ meV), which is similar to R-stacking bilayers formed at $\theta \sim 0^{\circ}$. In both devices, the coupling strength $J\sim 8$ meV is weaker than but of the same order of magnitude as aligned bilayers with $\theta$ close to $0^\circ$ or  $60^\circ$.

The revival of hybrid excitons in these two bilayers can be understood as a direct result of interlayer tunneling induced by a \moire lattice that is nearly commensurate with the monolayer lattices. The two twist angles are close to the commensurate angles $21.8^{\circ}$ and 38.2$^{\circ}$, respectively. At the commensurate angles, the \moire reciprocal lattice constant is $1/\sqrt{7}$ of the monolayer reciprocal lattice constant, and corners of the Brillouin zones of the two monolayers become connected by \moire reciprocal lattice vectors, as illustrated in Figs.~\ref{fig:coupling}(d) and ~\ref{fig:coupling}(e). That is, the \MoSe and \WS band minima overlap again in the \moire reciprocal lattice, allowing strong nearly-resonant tunneling between the intra- and inter-layer states.
Specifically, when $\theta \approx 21.8^{\circ}$, $K$-valley of \MoSe and $K'$-valley of \WS are connected by \moire reciprocal lattice vectors and are folded into equivalent momentum in the \moire Brillouin zone (Figs.~\ref{fig:coupling}(d)). The corresponding hybridized excitons have the same valley configuration as those in bilayers with $\theta\sim 60^{\circ}$, which is consistent with the observed negative detuning in both cases.  When $\theta \approx 38.2^{\circ}$, $K$-valley of \MoSe and $K$-valley of \WS are folded into equivalent momentum in the \moire Brillouin zone (Figs.~\ref{fig:coupling}(e)), and the corresponding hybridized excitons have the same valley configuration as those in bilayers with $\theta\sim 0^{\circ}$, consistent with the observed positive detuning in both cases. Moreover, since interlayer tunneling only needs one Umklapp scattering by a \moire reciprocal lattice vector, the tunneling strength remains of the same order of magnitude as in angularly aligned bilayers. Therefore, the strong revival of the hybrid excitons and their similarities with the angularly-aligned bilayers show again the critical role of \moire lattice in interlayer tunneling.

\begin{figure*}[t]
  \includegraphics[width=\linewidth]{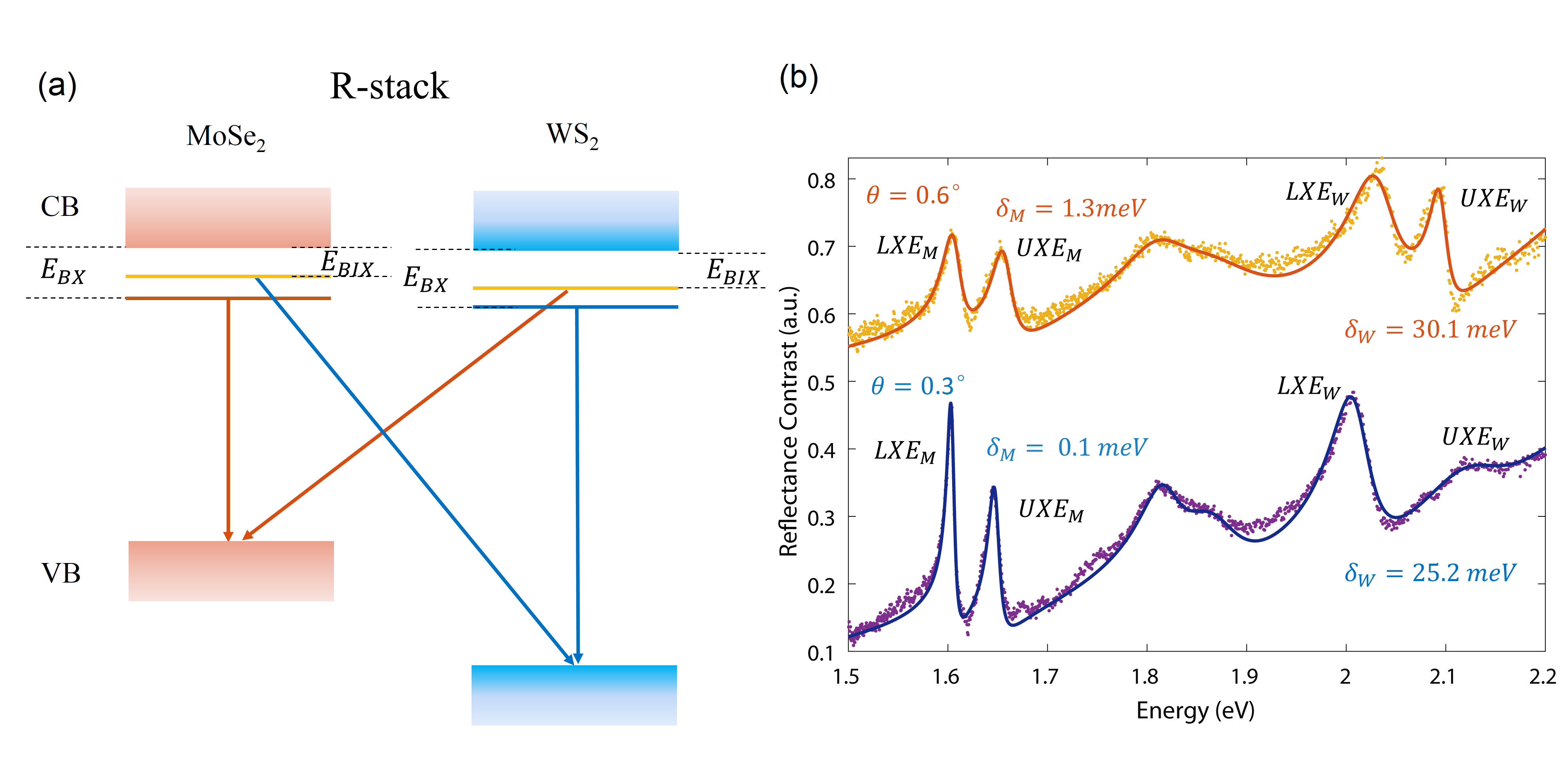}
	\caption{Comparison of $MoA$ and $WA$ hybrid states. (a) Band diagram of R-stacking \MoSe/\WS bilayers. The conduction and valence bands are represented by broad continuous bands. Exciton states are represented by the horizontal solid lines. Arrows represent spin-conserved exciton transitions, which are lowered in energy from the band-to-band transition by a binding energy. $E_{BX}$ and $E_{BIX}$ denote the binding energies for the intra- and inter-layer transitions, respectively. (c) RC spectra of both  $MoA$ and $WA$ hybrid excitons from bilayers with $\theta\sim 0^{\circ}$. Dots are data and solid lines are fits by transfer matrix calculations. The corresponding $\theta$ and detuning $\delta_M$ and $\delta_W$ obtained from fitting are labeled by each spectrum.
}
    \label{fig:BE}
\end{figure*}

In the above discussion, we have focused on hybrid states formed with the \MoSe A excitons, which feature large spectral weight, relatively narrow linewidths, and well-resolved doublets at small detunings. Hybrid states can also form with higher-energy bands, including the \MoSe B, \WS A and \WS B excitons. The B excitons have broader linewidths than the A excitons; as a result the doublets are not well resolved in most samples. The \WS A excitons have a broader linewidth than \MoSe A excitons and generally a larger detuning. We observe well-resolved $WA$ doublets only in bilayers with $\theta\sim 0^{\circ}$, corresponding to hybrid excitons formed by a hole in the \WS valence band and an electron tunneling between the \MoSe and \WS conduction bands (\cref{fig:BE}).

It is interesting to compare the detuning for $MoA$ and $WA$ states for $\theta\sim 0^{\circ}$, which we label as $\delta_{MoA}$ and $\delta_{WA}$, respectively. As shown in the schematic electronic band diagram in \cref{fig:BE}a, neglecting exciton binding energies, the detuning of the inter-layer transition from the intra-layer one is the same magnitude but opposite signs between the $MoA$ and $WA$ states. The sum of the two detuning should be zero. However, this is different from our observation that both the LHX states have larger spectral weight for both $MoA$ and $WA$ states. This can be understood as due to the weaker binding energy of interlayer excitons compared to intra-layer ones, resulting from electron-hole separation. The difference in intra- and inter-layer exciton binding energies, $\Delta E_B^R = E_{BX} - E_{BIX}$, adds to both $\delta_{MoA}$ and $\delta_{WA}$. Assuming $\Delta E_B^R$ is approximately the same for the $MoA$ and $WA$ state, the sum of $\delta_{MoA}$ and $\delta_{WA}$ becomes twice of $\Delta E_B$, or, $\Delta E_B^R = 1/2(\delta_{MoA} + \delta_{WA})$.  From our measurements of $MoA$ and $WA$ states in bilayers with $\theta < 1^\circ$, we estimate $\Delta E_B^R$ of $10$ to $16$~meV (\cref{fig:BE}b).
The value is significantly lower than predictions based on first principle calculations when interlayer tunneling is neglected \cite{latini_Interlayer_2017}. 

In summary, we demonstrate continuous tuning of hybrid \moire excitons that are formed by coupling between intralayer excitons and \moire lattice induced interlayer excitons. 
The twist-angle dependent hybridization of the excitons and their revival at the commensurate angles are direct manifestations of the discrete translational symmetry of the underlying \moire superlattice, which enables transitions that otherwise would not conserve momentum.
The freedom to tune the excitonic properties by tuning the \moire lattice provides a new venue to uncover fundamental properties of the system that are difficult to access otherwise, and may enable tuning and control of exotic states of matter
with novel applications in nanophotonics and quantum information science \cite{wu_Topological_2017,yu_Moire_2017,wu_theory_2018,wu_Hubbard_2018,tang_wse2/ws2_2019,regan_optical_2019,shimazaki_moire_2019}.

\section*{Methods}

\noindent\textbf{Sample fabrication.}
 Monolayer \MoSe, \WS and few layer hexagonal boron nitride (hBN) flakes are obtained by mechanical exfoliation from bulk crystals. A PET stamp was used to pick up the top hBN, \WS monolayer, \MoSe monolayer, and the bottom hBN under microscope. After picking up all the layers, PET stamp was then stamped onto sapphire substrate, and the PET was dissolved in dichloromethane for six hours at room temperature.

\noindent\textbf{Optical measurements.}
For low temperature measurements, the sample is kept in a 4~K cryostat (Montana Instrument).  The excitation and collection are carried out with a home-built confocal microscope with an objective lens with numerical aperture (NA) of 0.42. For reflection contrast measurement, white light from a tungsten halogen lamp is focused on the sample with a beam size of 10 $\mu$m in diameter. The spatial resolution is improved to be 2 $\mu$m by using pinhole combined with confocal lenses. 
The signal is detected using a Princeton Instruments spectrometer with a cooled charge-coupled camera.

\noindent\textbf{Coupled oscillator model on hybrid excitons.}
To extract the coupling strength $J$ and detuning $\delta$ of intralayer and interlayer excitons, we use the coupled oscillator model to describe the exciton hybridization, and write the Hamiltonian as:
    \[
    H=
    \begin{bmatrix}
    \label{eq:Rabi}
    E_{IX} & J  \\
    J & E_{X}
    \end{bmatrix}
    \]
    where $E_{IX}$ and $E_{X}$ represent the energies of uncoupled interlayer exciton and intralayer exciton, and $J$ is the tunneling strength of conduction bands electrons. By diagonalizing the matrix, the eigen energies can be extracted, and difference of the two eigen states is $\Delta E=\sqrt{4J^2+\delta^2}$, where $\delta=E_{IX}-E_{X}$. Since the oscillator strength of uncoupled interlayer exciton state is negligibly small compared with the intralayer exciton state, we set it as 0 in the calculation. The ratio of the oscillator strength between the hybrid excitons is $\frac{f_{LHX}}{f_{UHX}}=\frac{\sqrt{\delta^2+4J^2}+\delta}{\sqrt{\delta^2+4J^2}-\delta}$.

\noindent\textbf{Theory for hybrid moir\'e excitons.} We present a microscopic theory to account for our experimentally observed phenomena of hybrid moir\'e excitons. For definiteness, here we focus on $R$-stacking bilayers with a small twist angle $\theta$ near $0^{\circ}$, and the hybrid $MoA$ states at $K$ valley. The corresponding intralayer $A$ exciton state of MoSe$_2$ can be represented as
\begin{equation}
| \text{X} \rangle  = \frac{1}{\sqrt{\mathcal{A}}} \sum_{\boldsymbol{k}}
\psi_{\boldsymbol{k}} a_{c, M, (\boldsymbol{K}_{M}+\boldsymbol{k})}^{\dagger} a_{v, M, (\boldsymbol{K}_{M}+\boldsymbol{k})} | G \rangle,
\label{DX_Intra}
\end{equation}
where $| G \rangle$ represents the ground state of the system with fully filled valence bands, and $a_{c, M, (\boldsymbol{K}_{M}+\boldsymbol{k})}^{\dagger} a_{v, M, (\boldsymbol{K}_{M}+\boldsymbol{k})}$ creates a particle-hole excitation at valley $K$ in  MoSe$_2$ layer, and $\psi_{\boldsymbol{k}}$ is the relative-motion wave function. In Eq.~(\ref{DX_Intra}), $\mathcal{A}$ is the system area, $\boldsymbol{k}$ is the relative momentum between electron and holes, and $\boldsymbol{K}_{M}$ is the momentum of the $K$ point in the Brillouin zone of monolayer MoSe$_2$. The exciton state $| \text{X} \rangle$ in Eq.~(\ref{DX_Intra}) has a zero center-of-mass momentum, and therefore, can couple directly to the light and lead to optical absorption.

For interlayer excitons, we consider states with a generic finite center-of-mass momentum $\boldsymbol{Q}$
\begin{equation}
\begin{aligned}
| \boldsymbol{Q} \rangle _{\text{IX}} = \frac{1}{\sqrt{\mathcal{A}}} \sum_{\boldsymbol{k}}
\phi _{\boldsymbol{k}}  a_{c, W, (\boldsymbol{K}_W+\boldsymbol{k}+ \frac{m_{e,IX}}{M_{\text{IX}}}\boldsymbol{Q})}^{\dagger}
a_{v, M, (\boldsymbol{K}_M+\boldsymbol{k} - \frac{m_{h,IX}}{M_{\text{IX}}}\boldsymbol{Q})}  | G \rangle,
\end{aligned}
\label{IX_inter}
\end{equation}
where $a_{v, M, (\boldsymbol{K}_M+\boldsymbol{p})}$ and  $a_{c, W, (\boldsymbol{K}_W+\boldsymbol{p}')}^{\dagger}$ respectively create a hole in MoSe$_2$ valence band and an electron in WS$_2$ conduction band. $\phi_{\boldsymbol{k}}$ is the corresponding wave function, and  $\boldsymbol{k}$ is the relative momentum. $m_{e,IX}$ and  $m_{h,IX}$ are respectively electron and hole mass in the interlayer exciton, and $M_{\text{IX}}=m_{e,IX}+m_{h,IX}$ is the exciton total mass. The interlayer exciton state $| \boldsymbol{Q} \rangle _{\text{IX}}$ has an energy $\hbar \omega_{0}+\hbar^2  \boldsymbol{Q}^2/(2 M_{\text{IX}}) $, which includes an energy constant $\hbar \omega_{0}$ and  a kinetic energy of the center-of-mass motion. In Eq.~(\ref{IX_inter}), $\boldsymbol{K}_{W}$ is the momentum of the $K$ point in the Brillouin zone of monolayer WS$_2$, and differs from $\boldsymbol{K}_{M}$ due to lattice constant mismatch and misalignment.

The hybridization between intralayer and interlayer excitons is due to interlayer conduction-band tunneling \cite{wu_Topological_2017} in the moir\'e pattern, which is given in $+K$-valley by:
\begin{equation}
H_T= w \sum_{\boldsymbol{k}} \sum_{n=1,2,3} a_{c, W, (\boldsymbol{K}_W+\boldsymbol{k}+\boldsymbol{q}_n)}^{\dagger} a_{c, M, (\boldsymbol{K}_M+\boldsymbol{k})}+\text{H.c.},
\label{HTun}
\end{equation}
where $w$ is a tunneling parameter. In Eq.~(\ref{HTun}), $\boldsymbol{q}_1$, $\boldsymbol{q}_2$ and $\boldsymbol{q}_3$ are momenta that compensates the momentum shift between  $\boldsymbol{K}_W$ and $\boldsymbol{K}_M$ and are connected by the \moire reciprocal lattice vectors.
$\boldsymbol{q}_1$ is equal to $\boldsymbol{K}_M-\boldsymbol{K}_W$, while $\boldsymbol{q}_2$ and $\boldsymbol{q}_3$ are respectively related to $\boldsymbol{q}_1$ by $2\pi/3$ and $4\pi/3$ rotations. $|\boldsymbol{q}_1|$ is given by $4\pi/(3 a_M)$, and $a_M$ is the moir\'e period approximated by $a_0/\sqrt{\theta^2+\epsilon^2}$, where $a_0$ is the monolayer lattice constant, and $\epsilon$ is the lattice constant mismatch $|a_0-a_0'|/a_0$ between the two layers\cite{rasmussen_Computational_2015}.
In Eqs.~(\ref{DX_Intra}), (\ref{IX_inter}) and (\ref{HTun}), the spin label is not shown explicitly, and we consider spin up states.

This interlayer tunneling Hamiltonian $H_T$ hybridizes an intralayer exciton $| \text{X} \rangle $ with an interlayer exciton $| \text{IX} \rangle$, which shares the same angular momentum as $| \text{X} \rangle $ and can be written as \cite{wu_theory_2018}:
\begin{equation}
| \text{IX} \rangle = \frac{1}{\sqrt{3}} \Big(|\boldsymbol{q}_1\rangle_{\text{IX}}+|\boldsymbol{q}_2\rangle_{\text{IX}}+|\boldsymbol{q}_3\rangle_{\text{IX}} \Big).
\end{equation}
The energy difference $\delta$ between $| \text{IX} \rangle$ and $| \text{X} \rangle$ is
\begin{equation}
\begin{aligned}
\delta &= E_{IX}-E_{X}\\
& = \delta_0 + \frac{\hbar^2 \boldsymbol{q}_1^2}{2 M_{\text{IX}}} \\
& = \tilde{\delta}_0 + \Big(\frac{4\pi}{3 a_0}\Big)^2 \frac{\hbar^2 \theta^2}{2 M_{\text{IX}}},
\end{aligned}
\label{Ddetuning}
\end{equation}
where the energy $E_{intra}$ of intralayer exciton $| \text{X} \rangle $ is assumed to be independent of the twist angle $\theta$, while the energy $E_{inter}$ of interlayer exciton $| \text{IX} \rangle $ increases with increasing $\theta$ due to its kinetic energy. Equation~(\ref{Ddetuning}) provides a quantitative description of the experimentally observed $\theta$ dependence of the detuning $\delta$ when $\theta$ is small.

The coupling $J$ between $| \text{IX} \rangle$ and $| \text{X} \rangle$ due to the interlayer tunneling is
\begin{equation}
J= \langle \text{IX} | H_T | \text{X} \rangle = \frac{\sqrt{3} w} {\mathcal{A}} \sum_{\boldsymbol{k}} \phi_{\boldsymbol{k}+\frac{m_{h,IX}}{M_{\text{IX}}} \boldsymbol{q}_1}^* \psi_{\boldsymbol{k}},
\label{J_theory}
\end{equation}
where $\phi_{\boldsymbol{k}}$ and $\psi_{\boldsymbol{k}}$ are respectively the relative-motion wave function for interlayer and intralayer excitons with the normalizations $(1/\mathcal{A})\sum_{\boldsymbol{k}} |\psi_{\boldsymbol{k}}|^2=1$ and $(1/\mathcal{A})\sum_{\boldsymbol{k}} |\phi_{\boldsymbol{k}}|^2=1$. Because of the momentum shift $(m_{h,IX} / M_{\text{IX}}) \boldsymbol{q}_1$ in the integral of Eq.~(\ref{J_theory}),  $J$ decreases with increasing $\theta$, which agrees with the experimentally observed $\theta$ dependence of $J$. At $\theta=0^{\circ}$, $J$ can be approximated by $\sqrt{3} w$. Using our experimentally measured value of $J$, we can estimate the interlayer tunneling $w$ to be about 14 meV.

We make two remarks about the theory. (1) While $R$ stacking configuration and the MoA hybrid exctions are assumed in the above analysis, Equations~(\ref{Ddetuning}) and (\ref{J_theory}) apply equally well to $H$ stacking configuration and other hybrid excitons, but the exact parameter values can be different for different cases.  (2) Moir\'e pattern can in principle lead to additional bright exciton states besides those that are studied  above. Here we only consider hybrid exciton states made of $| \text{X} \rangle $ and $| \text{IX} \rangle $, because this type of hybrid states have the largest oscillator strengths. In summary, the theory presented here lays a microscopic foundation for the phenomenological coupled oscillator model, provides a microscopic explanation for the experimentally observed $\theta$ dependence of $J$ and $\delta$, and allows us to estimate the interlayer exciton total mass $M_{\text{IX}}$ and the interlayer tunneling $w$ from the optical spectra.

\noindent\textbf{Data availability} Data are available on request from the authors.
\noindent\textbf{Competing interests} The authors declare that they have no competing financial interests.
\noindent\textbf{Author Contributions}
  H.D., L.Z.conceived the experiment. L.Z. and Z.Z fabricated the device and performed the measurements. F.W. performed the modeling and calculations.  L.Z. and H.D. performed data analysis. R.G performed tunneling estimation. D.W, S.H, K.K. and T.G assisted the fabrication. K.W. and T.T grew hBN single crystals. H.D. and S.F. supervised the projects. L.Z, F.W. and H.D. wrote the paper with inputs from other authors. All authors discussed the results, data analysis and the paper.
\noindent\textbf{Acknowledgment}
L.Z.,R.G,S.H, S.F.and H.D.acknowledge the support by the Army Research Office under Awards W911NF-17-1-0312. F. W. is supported by Laboratory for Physical Sciences. K.W. and T.T. acknowledge support from the Elemental Strategy Initiative conducted by the MEXT, Japan and the CREST (JPMJCR15F3), JST. S.F. also acknowledges support by the U.S. Department of Energy, Office of Science, Office of Basic Energy Sciences, under Award Number DE-SC0017971.

\section*{References}
\bibliographystyle{naturemag}
\bibliography{Reference}

\begin{thebibliography}{10}
\expandafter\ifx\csname url\endcsname\relax
  \def\url#1{\texttt{#1}}\fi
\expandafter\ifx\csname urlprefix\endcsname\relax\def\urlprefix{URL }\fi
\providecommand{\bibinfo}[2]{#2}
\providecommand{\eprint}[2][]{\url{#2}}

\bibitem{geim_Van_2013}
\bibinfo{author}{Geim, A.~K.} \& \bibinfo{author}{Grigorieva, I.~V.}
\newblock \bibinfo{title}{Van der {Waals} heterostructures}.
\newblock \emph{\bibinfo{journal}{Nature}} \textbf{\bibinfo{volume}{499}},
  \bibinfo{pages}{419--425} (\bibinfo{year}{2013}).

\bibitem{novoselov_2D_2016}
\bibinfo{author}{Novoselov, K.~S.}, \bibinfo{author}{Mishchenko, A.},
  \bibinfo{author}{Carvalho, A.} \& \bibinfo{author}{Castro~Neto, A.~H.}
\newblock \bibinfo{title}{2d materials and van der {Waals} heterostructures}.
\newblock \emph{\bibinfo{journal}{Science}} \textbf{\bibinfo{volume}{353}},
  \bibinfo{pages}{aac9439--aac9439} (\bibinfo{year}{2016}).

\bibitem{rivera_valley-polarized_2016}
\bibinfo{author}{Rivera, P.} \emph{et~al.}
\newblock \bibinfo{title}{Valley-polarized exciton dynamics in a 2d
  semiconductor heterostructure}.
\newblock \emph{\bibinfo{journal}{Science}} \textbf{\bibinfo{volume}{351}},
  \bibinfo{pages}{688--691} (\bibinfo{year}{2016}).

\bibitem{kuwabara_Anomalous_1990}
\bibinfo{author}{Kuwabara, M.}, \bibinfo{author}{Clarke, D.~R.} \&
  \bibinfo{author}{Smith, D.~A.}
\newblock \bibinfo{title}{Anomalous superperiodicity in scanning tunneling
  microscope images of graphite}.
\newblock \emph{\bibinfo{journal}{Applied Physics Letters}}
  \textbf{\bibinfo{volume}{56}}, \bibinfo{pages}{2396--2398}
  (\bibinfo{year}{1990}).

\bibitem{zhang_interlayer_2017}
\bibinfo{author}{Zhang, C.} \emph{et~al.}
\newblock \bibinfo{title}{Interlayer couplings, moir\'e patterns, and
  2$\uppercase{D}$ electronic superlattices in {MoS}$_{2}$/{WSe}$_{2}$
  hetero-bilayers}.
\newblock \emph{\bibinfo{journal}{Science advances}}
  \textbf{\bibinfo{volume}{3}}, \bibinfo{pages}{e1601459}
  (\bibinfo{year}{2017}).

\bibitem{cao_Unconventional_2018}
\bibinfo{author}{Cao, Y.} \emph{et~al.}
\newblock \bibinfo{title}{Unconventional superconductivity in magic-angle
  graphene superlattices}.
\newblock \emph{\bibinfo{journal}{Nature}} \textbf{\bibinfo{volume}{556}},
  \bibinfo{pages}{43--50} (\bibinfo{year}{2018}).

\bibitem{cao_correlated_2018}
\bibinfo{author}{Cao, Y.} \emph{et~al.}
\newblock \bibinfo{title}{Correlated insulator behaviour at half-filling in
  magic-angle graphene superlattices}.
\newblock \emph{\bibinfo{journal}{Nature}} \textbf{\bibinfo{volume}{556}},
  \bibinfo{pages}{80--84} (\bibinfo{year}{2018}).

\bibitem{chen_Signatures_2019}
\bibinfo{author}{Chen, G.} \emph{et~al.}
\newblock \bibinfo{title}{Signatures of tunable superconductivity in a trilayer
  graphene moir\'e superlattice}.
\newblock \emph{\bibinfo{journal}{Nature}} \textbf{\bibinfo{volume}{572}},
  \bibinfo{pages}{215--219} (\bibinfo{year}{2019}).

\bibitem{chen_Evidence_2019}
\bibinfo{author}{Chen, G.} \emph{et~al.}
\newblock \bibinfo{title}{Evidence of a gate-tunable {Mott} insulator in a
  trilayer graphene moir\'e superlattice}.
\newblock \emph{\bibinfo{journal}{Nature Physics}}
  \textbf{\bibinfo{volume}{15}}, \bibinfo{pages}{237--241}
  (\bibinfo{year}{2019}).

\bibitem{wu_Topological_2017}
\bibinfo{author}{Wu, F.}, \bibinfo{author}{Lovorn, T.} \&
  \bibinfo{author}{MacDonald, A.~H.}
\newblock \bibinfo{title}{Topological exciton bands in moir\'e
  heterojunctions}.
\newblock \emph{\bibinfo{journal}{Phys. Rev. Lett.}}
  \textbf{\bibinfo{volume}{118}}, \bibinfo{pages}{147401}
  (\bibinfo{year}{2017}).

\bibitem{yu_Moire_2017}
\bibinfo{author}{Yu, H.}, \bibinfo{author}{Liu, G.-B.}, \bibinfo{author}{Tang,
  J.}, \bibinfo{author}{Xu, X.} \& \bibinfo{author}{Yao, W.}
\newblock \bibinfo{title}{Moir\'e excitons: {From} programmable quantum emitter
  arrays to spin-orbit–coupled artificial lattices}.
\newblock \emph{\bibinfo{journal}{Science Advances}}
  \textbf{\bibinfo{volume}{3}}, \bibinfo{pages}{e1701696}
  (\bibinfo{year}{2017}).

\bibitem{wu_theory_2018}
\bibinfo{author}{Wu, F.}, \bibinfo{author}{Lovorn, T.} \&
  \bibinfo{author}{MacDonald, A.~H.}
\newblock \bibinfo{title}{Theory of optical absorption by interlayer excitons
  in transition metal dichalcogenide heterobilayers}.
\newblock \emph{\bibinfo{journal}{Phys. Rev. B}} \textbf{\bibinfo{volume}{97}},
  \bibinfo{pages}{035306} (\bibinfo{year}{2018}).

\bibitem{wu_Hubbard_2018}
\bibinfo{author}{Wu, F.}, \bibinfo{author}{Lovorn, T.}, \bibinfo{author}{Tutuc,
  E.} \& \bibinfo{author}{MacDonald, A.}
\newblock \bibinfo{title}{Hubbard {Model} {Physics} in {Transition} {Metal}
  {Dichalcogenide} {Moir\'e} {Bands}}.
\newblock \emph{\bibinfo{journal}{Phys. Rev. Lett.}}
  \textbf{\bibinfo{volume}{121}}, \bibinfo{pages}{026402}
  (\bibinfo{year}{2018}).

\bibitem{jin_Observation_2019}
\bibinfo{author}{Jin, C.} \emph{et~al.}
\newblock \bibinfo{title}{Observation of moir\'e excitons in
  {WSe}$_{2}$/{WS}$_{2}$ heterostructure superlattices}.
\newblock \emph{\bibinfo{journal}{Nature}} \textbf{\bibinfo{volume}{567}},
  \bibinfo{pages}{76--80} (\bibinfo{year}{2019}).

\bibitem{tran_Evidence_2019}
\bibinfo{author}{Tran, K.} \emph{et~al.}
\newblock \bibinfo{title}{Evidence for moir\'e excitons in van der {Waals}
  heterostructures}.
\newblock \emph{\bibinfo{journal}{Nature}} \textbf{\bibinfo{volume}{567}},
  \bibinfo{pages}{71--75} (\bibinfo{year}{2019}).

\bibitem{seyler_Signatures_2019}
\bibinfo{author}{Seyler, K.~L.} \emph{et~al.}
\newblock \bibinfo{title}{Signatures of moir\'e-trapped valley excitons in
  {MoSe}$_{2}$/{WSe}$_{2}$ heterobilayers}.
\newblock \emph{\bibinfo{journal}{Nature}} \textbf{\bibinfo{volume}{567}},
  \bibinfo{pages}{66--70} (\bibinfo{year}{2019}).

\bibitem{wang_Evidence_2019}
\bibinfo{author}{Wang, Z.} \emph{et~al.}
\newblock \bibinfo{title}{Evidence of high-temperature exciton condensation in
  two-dimensional atomic double layers}.
\newblock \emph{\bibinfo{journal}{Nature}} \textbf{\bibinfo{volume}{574}},
  \bibinfo{pages}{76--80} (\bibinfo{year}{2019}).

\bibitem{tang_wse2/ws2_2019}
\bibinfo{author}{Tang, Y.} \emph{et~al.}
\newblock \bibinfo{title}{{WSe}$_2$/{WS}$_2$ moir\'e superlattices: a new
  {Hubbard} model simulator}.
\newblock \emph{\bibinfo{journal}{arXiv:1910.08673 [cond-mat]}}
  (\bibinfo{year}{2019}).

\bibitem{regan_optical_2019}
\bibinfo{author}{Regan, E.~C.} \emph{et~al.}
\newblock \bibinfo{title}{Optical detection of {Mott} and generalized {Wigner}
  crystal states in {WSe}$_2$/{WS}$_2$ moir\'e superlattices}.
\newblock \emph{\bibinfo{journal}{arXiv:1910.09047 [cond-mat]}}
  (\bibinfo{year}{2019}).

\bibitem{shimazaki_moire_2019}
\bibinfo{author}{Shimazaki, Y.} \emph{et~al.}
\newblock \bibinfo{title}{Moir\'e superlattice in a {MoSe}$_2$/{hBN}/{MoSe}$_2$
  heterostructure: from coherent coupling of inter- and intra-layer excitons to
  correlated {Mott}-like states of electrons}.
\newblock \emph{\bibinfo{journal}{arXiv:1910.13322 [cond-mat]}}
  (\bibinfo{year}{2019}).

\bibitem{alexeev_Resonantly_2019}
\bibinfo{author}{Alexeev, E.~M.} \emph{et~al.}
\newblock \bibinfo{title}{Resonantly hybridized excitons in moir\'e
  superlattices in van der {Waals} heterostructures}.
\newblock \emph{\bibinfo{journal}{Nature}} \textbf{\bibinfo{volume}{567}},
  \bibinfo{pages}{81--86} (\bibinfo{year}{2019}).

\bibitem{latini_Interlayer_2017}
\bibinfo{author}{Latini, S.}, \bibinfo{author}{Winther, K.~T.},
  \bibinfo{author}{Olsen, T.} \& \bibinfo{author}{Thygesen, K.~S.}
\newblock \bibinfo{title}{Interlayer {Excitons} and {Band} {Alignment} in
  {MoS}2/{hBN}/{WSe}2 van der {Waals} {Heterostructures}}.
\newblock \emph{\bibinfo{journal}{Nano Lett.}} \textbf{\bibinfo{volume}{17}},
  \bibinfo{pages}{938--945} (\bibinfo{year}{2017}).

\bibitem{hsu_Tailoring_a}
\bibinfo{author}{Hsu, W.-T.} \emph{et~al.}
\newblock \bibinfo{title}{Tailoring excitonic states of van der {Waals}
  bilayers through stacking configuration, band alignment and valley-spin}.
\newblock \emph{\bibinfo{journal}{arXiv:1903.02157 [cond-mat]}}
  (\bibinfo{year}{2019}).

\bibitem{li_Measurement_2014}
\bibinfo{author}{Li, Y.} \emph{et~al.}
\newblock \bibinfo{title}{Measurement of the optical dielectric function of
  monolayer transition-metal dichalcogenides: {MoS}$_{2}$,
  $\mathrm{Mo}\mathrm{S}{\mathrm{e}}_{2}$, {WS}$_{2}$, and
  $\mathrm{WS}{\mathrm{e}}_{2}$}.
\newblock \emph{\bibinfo{journal}{Phys. Rev. B}} \textbf{\bibinfo{volume}{90}},
  \bibinfo{pages}{205422} (\bibinfo{year}{2014}).

\bibitem{Note1}
\bibinfo{note}{Considering the lattice mismatch between the two layers (see
  Methods) would add a small correction to the measured spin-orbit splitting.
  Using Eq.~1 and the fitted effective masses, we obtain a value of 26.4 meV
  for spin-orbit coupling based on the measurement.}

\bibitem{liu_Electronic_2015}
\bibinfo{author}{Liu, G.-B.}, \bibinfo{author}{Xiao, D.}, \bibinfo{author}{Yao,
  Y.}, \bibinfo{author}{Xu, X.} \& \bibinfo{author}{Yao, W.}
\newblock \bibinfo{title}{Electronic structures and theoretical modelling of
  two-dimensional group-{VIB} transition metal dichalcogenides}.
\newblock \emph{\bibinfo{journal}{Chem. Soc. Rev.}}
  \textbf{\bibinfo{volume}{44}}, \bibinfo{pages}{2643--2663}
  (\bibinfo{year}{2015}).

\bibitem{rasmussen_Computational_2015}
\bibinfo{author}{Rasmussen, F.~A.} \& \bibinfo{author}{Thygesen, K.~S.}
\newblock \bibinfo{title}{Computational 2d {Materials} {Database}: {Electronic}
  {Structure} of {Transition}-{Metal} {Dichalcogenides} and {Oxides}}.
\newblock \emph{\bibinfo{journal}{J. Phys. Chem. C}}
  \textbf{\bibinfo{volume}{119}}, \bibinfo{pages}{13169--13183}
  (\bibinfo{year}{2015}).

\end{thebibliography}


\end{document}